\documentclass[article]{epl}

\usepackage{amsmath}
\usepackage{amsfonts}
\usepackage{amssymb}

\usepackage{color}

\title{The Casimir Effect for the Bose-Gas in Slabs}
\author{Philippe A. Martin\thanks{E-mail: \email{Philippe-Andre.Martin@epfl.ch}}\inst{1}
\and Valentin A. Zagrebnov\thanks{E-mail: \email{Valentin.Zagrebnov@cpt.univ-mrs.fr}} \inst{2,3}}

\institute{
  \inst{1} Institute of Theoretical Physics - Swiss Federal Institute of Technology Lausanne,
  CH-1015 Lausanne EPFL, Switzerland\\
  \inst{2} Universit\'{e} de la M\'{e}diterran\'{e}e
(Aix-Marseille II), Facult\'{e} des Sciences de Luminy\\
\inst{3} Centre de Physique Th\'{e}orique, Luminy-Case 907, 13288
Marseille Cedex 9, France }
\pacs{05.30.-d}{Quantum statistical
mechanics} \pacs{05.30.Jp}{Boson systems}
\pacs{03.75.Hh}{Static
properties of condensates; thermodynamical,statistical and
structural properties}
\begin{document}

\maketitle

\begin{abstract}
We study the Casimir effect for the perfect  Bose-gase in the slab geometry
for various boundary conditions. We show that the grand canonical potential per unit area
at the bulk critical chemical potential $\mu=0$ has the standard asymptotic
form with universal Casimir terms.
\end{abstract}

In contrast to the well-known \textit{Casimir effect} for the photon
gas, see e.g.\cite{Milonni}, this effect for the massive quantum
particles (\textit{quantum gases}) is much less explored. In the
present letter we consider the case of the perfect Bose-gas. The
perfect Bose-gas is the simplest quantum mechanical system showing the
spontaneous breaking of a continuous symmetry (Bose-Einstein
condensation) for which the Casimir effect should appear. To our
knowledge, this has surprisingly not been studied before.

We first recall the definition of the \textit{Casimir amplitude} at
critical points or for phases with broken symmetry
\cite{Brankov}\cite{Krech}. Let
\begin{equation}\label{slab-poten}
\varphi_d(T)= \lim_{L\rightarrow\infty}\frac{1}{L^2}\Phi_L(T,d)
\end{equation}
be a limiting thermodynamic potential per unit area  in the
\textit{3-dimensional slab geometry} $\infty \times \infty \times
d$, where $\Phi_L(T,d)$ is the finite-volume thermodynamic
potential in the box $\Lambda = L \times L \times d$ and $T$ denotes the
temperature. The finite size scaling analysis in the
\textit{critical regime} shows that large
$d$-asymptotics of (\ref{slab-poten}) has usually the form:
\begin{equation}\label{crit-asymp}
\varphi_d(T)= d \, \varphi_{bulk}(T) + \varphi_{surf}(T) +
\frac{\Delta (T)}{d^2} + \ldots   \,.
\end{equation}
Here $\varphi_{bulk}(T)$ is the potential density in thermodynamic
limit and $\varphi_{surf}(T)$ is the surface potential correction.
The coefficient $\Delta (T)$ in the third term in (\ref{crit-asymp})
is called the \textit{Casimir amplitude}. The \textit{Casimir force}
between the slab faces is defined as
\begin{equation}\label{Cas-force}
F(d) = - \,\partial_d [\varphi_d(T)- d \, \varphi_{bulk}(T)] = 2 \
\frac{\Delta (T)}{d^3} + \ldots \,.
\end{equation}
In this regime, i.e. at a critical point or in phases with a long-range correlations
generated by the broken symmetry, the value of $\Delta (T)$ is
expected to be non-zero and \textit{universal}, depending only on
the system and the boundary condition universality classes. Out of
the \textit{critical regime} the finite size corrections in slab are
expected to be exponentially small, thus the Casimir amplitude
$\Delta (T) = 0$.

The perfect Bose-gas \textit{grand-canonical thermodynamic
potential} in the box $L \times L \times d$ has the form
\cite{Huang}
\begin{equation}\label{g-c-potent}
\Phi_{L,d} (T,\mu) = \beta^{-1} \sum_{\textbf{k}} \ln \left\{ 1 -
e^{-\beta (\varepsilon(\textbf{k})- \mu)}\right\}\,, \ \ \, \ \
\varepsilon(\textbf{k})= \frac{\hbar^2}{2 m} (k^{2}_{x} + k^{2}_{y}
+ k^{2}_{z}) \,\,,
\end{equation}
where $\beta = (k_B T)^{-1}$ and $\mu$ is the chemical potential.
The sum in (\ref{g-c-potent}) runs over the set defined by boundary
conditions (b.c.) for the Laplacian operator in the box $L \times L
\times d$. Below we shall consider (for simplicity) the periodic
b.c. in the $x$-$y$ directions:
\begin{equation}\label{XY-b-c}
k_{x} = \frac{2\pi}{L} n_{x}\,, \quad k_{y} = \frac{2\pi}{L} n_{y}\ \ \ \,
\ \ \ n_{x},n_{y}= 0, \pm 1,
\pm 2, \ldots \,,
\end{equation}
and three cases of b.c., Dirichlet ($D$), Neumann ($N$) and periodic ($P$), in the $z$ direction:
\begin{eqnarray}\label{D-N-b-c-Z-dir}
(D) \ \ \ k_{z} = \frac{\pi}{d} n_{z} &,&  n_{z}= 1, 2, \ldots \ ; \
\ (N) \ \ \ k_{z} = \frac{\pi}{d} n_{z} \ , \ n_{z}= 0, 1, 2, \ldots \ \ \ \\
&& (P) \ \ \ k_{z} = \frac{2\pi}{d} n_{z} \ , \ n_{z}= 0, \pm 1, \pm
2, \ldots  \, . \label{P-b-c-Z-dir}
\end{eqnarray}
Therefore, in the \textit{slab limit}, $L \rightarrow \infty$, the grand-canonical thermodynamic
potential per unit area is
\begin{eqnarray}
\varphi_d (T,\mu)&=& \lim_{L \rightarrow \infty}\frac{1}{L^2} \,\, \Phi_{L,d} (T,\mu)
= \frac{1}{\beta(2\pi)^{2}}\int_{\mathbb{R}^{2}}d^2\mathbf{q}\,\,
\sum_{k_z} \ln \left\{1 -
e^{-\beta\left[\varepsilon(\mathbf{q})+ \varepsilon(k_{z})
- \mu\right]} \right\}
\nonumber\\ &=& -\frac{1}{(2\pi)^{2}}\int_{\mathbb{R}^{2}}d^2\mathbf{q}\,\,\sum_{k_z}
\frac{\varepsilon(\mathbf{q})}{{e^{\beta\left[\varepsilon(\mathbf{q})+
\varepsilon(k_{z}) - \mu\right]}}- 1}\,,
\label{pot-unit-area}
\end{eqnarray}
where $\mathbf{q}=(q_{x},q_{y})$ is a two dimensional wave-vector in the $(x,y)$ plane,
$\varepsilon(\mathbf{q})= \hbar^2 (q^{2}_x + q^{2}_y)/2m$ and
$\varepsilon(k_{z})= \hbar^2 k^{2}_z/2m$. The last equality results from an integration by
part with respect to $q=|\mathbf{q}|$.

It is well-known that in the \textit{bulk limit}, $L \rightarrow
\infty \,,\,d \rightarrow \infty$, the Bose-Einstein condensation
for the perfect Bose-gas may occur only for $\mu = 0$. Therefore in
the sequel we distinguish the two regimes : $\mu < 0$
(\textit{normal phase}) and $\mu = 0$ (\textit{condensed phase}).
Our results for the asymptotic expressions of $\varphi_d (T,\mu)$ as
$d \rightarrow \infty$ are the following:\\

(a) If $\mu < 0$, then
\begin{equation}\label{asympt-mu<0}
\varphi_d (T,\mu) = -d \, p_{bulk}(T,\mu) + \varphi_{surf}(T,\mu) + O
\left(e^{-Const\cdot\sqrt{-\beta\mu}\,d/\lambda}\right) ,
\end{equation}
where
\begin{equation}\label{press}
p_{bulk}(T,\mu)= -\frac{1}{\beta (2\pi)^3}\int d^3\mathbf{k}
\ln\left(1 - e^{-\beta\left[\varepsilon(\mathbf{k}) -
\mu\right]}\right)
\end{equation}
is the standard Bose-gas pressure and $\lambda = \hbar
\sqrt{\beta/m}$ denotes the \textit{thermal wave-length}. We get
that $\varphi^{(P)}_{surf}(T,\mu)= 0$ for periodic b.c., but
\begin{equation}\label{surf-D-N}
\varphi^{(D,N)}_{surf}(T,\mu)= \pm \frac{1}{2(2\pi)^2}\int
d^2\mathbf{q}
\frac{\varepsilon(\mathbf{q})}{{e^{\beta\left[\varepsilon(\mathbf{q})-
\mu\right]}}- 1}
\end{equation}
with $+$ for Dirichlet and $-$ for Neumann b.c.\\

(b) If $\mu = 0$, then we obtain for the periodic b.c. the
asymptotics
\begin{eqnarray}
\varphi_d (T,0) &=& -d \,\, p_{bulk}(T,0) +\frac{k_{B}T}{d^{2}}\left[
\Delta^{(P)}(T) +
O\left(({\lambda}/{d})^{M}\right)\right] ,  \quad\mbox{for any} \,\,\, M\geq 1 \nonumber\\
\Delta^{(P)}(T)&=& -\frac{k_B T \,
\zeta (3)}{\pi}
\label{asympt-mu=0-Per}
\end{eqnarray}
where $\zeta(s)=\sum_{n=1}^{\infty} {n^{-s}}$ is the \textit{Riemann
zeta-function}. For the Dirichlet and Neumann b.c. we find
\begin{eqnarray}
&&\varphi_d (T,0) = -d\,\, p_{bulk}(T,0) +
\varphi^{(D,N)}_{surf}(T,0) + \frac{k_{B}T}{d^{2}}\left[
\Delta^{(D,N)}(T) + O\left(({\lambda}/{d})^{M}\right)\right] ,\,\,
\mbox{for any} \,\, M\geq 1 \nonumber\\
&&\Delta^{(D,N)}(T)= -\frac{k_B T \, \zeta (3)}{8 \pi}\,,
\label{asympt-mu=0-D-N}
\end{eqnarray}
 and $\varphi^{(D,N)}_{surf}(T,0)$ is given by (\ref{surf-D-N}).\\

We see that for Dirichlet and Neumann b.c. the Casimir amplitudes (\ref{asympt-mu=0-D-N})
are the same, implying identical \textit{attractive} forces in slab
in presence of the Bose-Einstein condensation.\\

To obtain  these results we consider separately the cases $\mu<0$ and $\mu=0$.

For $\mu < 0$ and Dirichlet b.c. (see (\ref{D-N-b-c-Z-dir})) the
result (a) follows from representation of (\ref{pot-unit-area}) as
the sum:
\begin{equation}\label{discr-sum}
\varphi_d (T,\mu)  = \sum_{n=1}^{\infty} \phi(d^{-1}
n)\,\,\,,\,\,\,\, \phi(u) =
-\frac{1}{(2\pi)^{2}}\int_{\mathbb{R}^{2}}d^2\mathbf{q}\,
\frac{\varepsilon(\mathbf{q})}{{e^{\left[\beta\varepsilon(\mathbf{q})+
(\pi \lambda u)^2/2 - \beta\mu\right]}}- 1} \,\,.
\end{equation}
 Since for $\mu < 0$ the function $\phi(u)$ is
infinitely differentiable at $u=0$ and all its derivatives vanish at
infinity, we can use the \textit{Euler-MacLaurin theorem} \cite{Euler}, \cite{chatter}
in the form
\begin{equation}\label{Euler}
\sum_{n=1}^{\infty} F(n) = \int_{0}^{\infty} ds \,F(s) - \frac{1}{2}
F(0) - \frac{B_2}{2!} F^{(1)}(0) - \frac{B_4}{4!} F^{(3)}(0) -
\ldots   \,,
\end{equation}
where $\left\{B_{2n}\right\}_{n=1}^{\infty}$ are the Bernoulli
numbers and $\left\{F^{(2n-1)}(s)\right\}_{n=1}^{\infty}$ denote the
corresponding  derivatives. When applying (\ref{Euler}) to
(\ref{discr-sum}), the two first terms of (\ref{Euler}) yield
respectively the bulk and surface contributions in
(\ref{asympt-mu<0}). Moreover, since $\phi(u)=\phi(-u)$ all odd
derivatives $\left\{\phi^{(2n-1)}(u)\right\}_{n=1}^{\infty}$ vanish
at $u=0$. So for large $d$ the corrections to the first two
terms are smaller than any power of $d^{\,-1}$. In fact, a more
elaborated estimate gives an exponentially small correction $O(e^{-Const\cdot
\,\sqrt{-\beta\mu}\,d/\lambda})$, see (\ref{estim-negat-mu}) below.\\

If $\mu = 0$, all derivatives $\phi^{(2n-1)}(u)$ of order $2n-1 > 2$
diverge at $u=0$. So instead of the Euler-MacLaurin formula one has
to use a more refined representation for the sum (\ref{discr-sum}).
To this end we use the \textit{Jacobi identity} (see e.g. \cite{Jacobi}, Ch.11):
\begin{equation}\label{Jacobi}
\sum_{n=1}^{\infty}e^{- \pi\, n^2 \, a} = \left(\frac{1}{2\sqrt{a}} -
\frac{1}{2}\right) + \frac{1}{\sqrt{a}}\sum_{n=1}^{\infty}e^{-
\pi\,n^2/a}\,\,\,,\,\,\,\,\,a > 0 \,\,,
\end{equation}
which can be obtained as a simple consequence of the Poisson
summation formula. First we represent (\ref{discr-sum}) by its low
activity series for $\mu < 0$ (setting
$v=\beta\epsilon(\mathbf{q})=\lambda^{2}q^{2}/2$)
\begin{eqnarray}\label{low-activ}
&&\varphi_d (T,\mu)= -\frac{1}{2\pi\beta\lambda^2}\int_{0}^{\infty}
dv \,v \sum_{n=1}^{\infty} \sum_{r=1}^{\infty}\, e^{\beta r \mu}
e^{- r [v + (\pi \lambda n/d)^2/2]} = \\ \nonumber
&&-\frac{1}{2\pi\beta\lambda^2}
\sum_{r=1}^{\infty}\, \frac{e^{\beta r \mu}}{r^2}
\sum_{n=1}^{\infty} e^{- \pi n^2 (r \pi (\lambda /d)^2/2)}
= -\frac{d}{\beta (\sqrt{2\pi}\lambda)^3} \sum_{r=1}^{\infty}\,
\frac{e^{\beta r \mu}}{r^{5/2}} + \frac{1}{4\pi \beta \lambda^2}
\sum_{r=1}^{\infty}\, \frac{e^{\beta r \mu}}{r^{2}} + \\
&& -\frac{2
d}{\beta (\sqrt{2\pi}\lambda)^3} \sum_{n=1}^{\infty}\sum_{r=1}^{\infty}\,
\frac{e^{\beta r \mu}}{r^{5/2}} e^{-2
\,(nd/\lambda)^2/r} \,,\nonumber
\end{eqnarray}
where the last equation follows from the the Jacobi identity (\ref{Jacobi})
with $a=r \pi (\lambda /d)^2/2$. These series are absolutely
convergent when $\mu=0$ and one recognises that the first term in
the right-hand side of (\ref{low-activ}) in nothing but $-d \,\,
p_{bulk}(T,0)$ and the second one coincides with
$\varphi^{(D)}_{surf}(T,0)$. To obtain for $\mu=0$ the large $d$
asymptotics of the last term we write the sum over $r$ in the form
$(\lambda/nd)^{3}\delta_n(d)S_n(d)$, where
\begin{equation}\label{sum-r}
S_n(d) = \sum_{r=1}^{\infty} \psi (\delta_n(d)r)
\,\,,\,\,\,\,\,\,\,\, \psi(x)= \frac{e^{-2 /x}}{x^{5/2}}
\,\,,\,\,\,\,\,\,\,\,\, \delta_n(d)= \left(\frac{\lambda}{n d}\right)^2 \,\,.
\end{equation}
Since the function
$\psi(x)$ is infinitely differentiable for $x\geq 0$ and all its
derivatives vanish at $x=0$ and at infinity, the Euler-MacLaurin formula (\ref{Euler})
yields for large
$d$ and $n$
\begin{equation}\label{E-M-sum-r}
\delta_n(d)S_n(d)= \int_{0}^{\infty}dx \psi (x) +
O\left((\delta_n(d))^M\right)= \frac{\sqrt{\pi}}{2^{5/2}} + O\left((\delta_n(d))^M\right)\,\,
\,\,\mbox{for any} \,\,\,\, M\geq 1 \,\,,
\end{equation}
i.e. the corrections to the integral are smaller than any power of $\delta_n(d)$.
Inserting (\ref{E-M-sum-r}) in the last term
of the right-hand side of (\ref{low-activ}) we get
\begin{equation}\label{last-term}
\frac{2d}{\beta (\sqrt{2\pi}\lambda)^3}\sum_{n=1}^{\infty} \sum_{r=1}^{\infty}\,
\frac{1}{r^{5/2}} e^{-2
\,(nd/\lambda)^2/r} = \frac{1}{8 \pi \beta d^2}\left[\sum_{n=1}^{\infty}\frac{1}{n^3} +
O\left(({\lambda}/{d})^{M}\right)\right]\,,\,\,\mbox{for any} \,\,\,\, M\geq 1 \,\,,
\end{equation}
that proves (\ref{asympt-mu=0-D-N}) for the Dirichlet b.c.
By (\ref{D-N-b-c-Z-dir}) for Neumann b.c. sums in (\ref{Jacobi}) and (\ref{low-activ})
start with $n=0$,
which implies changing the sign of the surface term (\ref{surf-D-N})
and gives (\ref{asympt-mu=0-D-N}) for Neumann b.c.

For periodic b.c. (\ref{P-b-c-Z-dir}) we have to use instead of  (\ref{discr-sum})
the representation
\begin{equation*}
\varphi_d (T,\mu)  = \sum_{n=-\infty}^{\infty} \phi(({d}/{2})^{-1}
n) = 2 \, \sum_{n=1}^{\infty} \phi(({d}/{2})^{-1}
n) + \phi(0)\,.
\end{equation*}
By consequence, the contribution of the $n=0$ term, which corresponds to the surface corrections,
will disappear and we obtain (\ref{asympt-mu=0-Per}).

The Jacobi identity allows also to prove the asymptotic (\ref{asympt-mu<0}). To this end notice
that double-sum in the last term of (\ref{low-activ}) has for $\mu < 0$ the estimate
\begin{eqnarray}\label{estim-negat-mu}
\sum_{r=1}^{\infty}\,
\frac{e^{\beta r \mu}}{r^{5/2}}\sum_{n=1}^{\infty} e^{-2
\,(nd/\lambda)^2/r}&\leq&  \sum_{r=1}^{\infty}\,
\frac{1}{r^{5/2}}\sum_{n=1}^{\infty} e^{\max_{r\geq1}(\beta r \mu -2
\,(nd/\lambda)^2/r)} \\ \nonumber
&=& \frac{\zeta(5/2)}{e^{2\sqrt{-2\beta \mu}d/\lambda}-1} = O(e^{-\sqrt{- 8\beta \mu}\,d/\lambda}),
\end{eqnarray}
that proves (\ref{asympt-mu<0}) for large $d$.\\

\noindent Now few remarks and comments are in order.

The grand potential of a free Fermion gas does not have a Casimir
term for any value of the chemical potential.  Indeed replacing the
Bose by the Fermi distribution in (\ref{pot-unit-area}) gives a
corresponding $\phi(u)$ function (\ref{discr-sum}) that is
infinitely differentiable at $u=0$ for all $\mu$. Since the function
and all its derivatives vanish at $u=0$ the Euler-MacLaurin formula
yields corrections smaller than any inverse power of $d$.

According to common wisdom, the Casimir force is due to Goldstone modes
that will occur in the bulk limit when a continuous symmetry is
spontaneously broken. Usually they are taken \textit{ad hoc} as a
part of phenomenological models, see e.g. \cite{ZRK} for the case of
superfluid films. In the free Bose-gas it is explicit: the excitations
$\varepsilon({\bf k})- \mu$ become gapless when $\mu=0$. Then by the
London-Placzek formula \cite{Zif},\cite{LP}, this generates long-range
particle-particle correlations
\begin{equation}\label{Long-Range-BEC}
\rho^{(2)}({\bf r}_{1},{\bf
r}_{2})-\rho^{2}\sim  \rho_{0}(T) |{\bf r}_{1}- {\bf r}_{2}|^{-1} ,
\end{equation}
as $|{\bf r}_{1}-{\bf r}_{2}|\to\infty$ in the condensed phase.
Here $\rho_{0}$ denotes the Bose-Einstein condensation
density of the perfect Bose-gas. Casimir forces are usually attributed to such
correlations in the critical regime. One can check that it is true
for the perfect Bose-gas because of long-range
correlations that appear in the slab limit.
\\

It is known that the perfect Bose gas particle number fluctuations
in the condensed phase are abnormal, i. e. proportional to volume
$V$ in contrast to the normal thermodynamical fluctuations
$\sqrt{V}$,  because of lack of superstability, see e.g. \cite{PZ}.
So we would like to warn that the results for Casimir amplitudes
might be very different from (\ref{asympt-mu=0-Per}),
(\ref{asympt-mu=0-D-N})
in the presence of a superstable interaction.\\

In the case of electromagnetic interactions, the Casimir term is
always present as a result of the long range of the forces. In the
standard calculation of the zero-temperature Casimir force between
perfect conductors \cite{Milonni} the Casimir term appears because
the third order derivative occuring in the appropriate
Euler-MacLaurin expansion does not vanish, contrary to the case (a)
above. This is due to the linear form $\hbar
\omega_{\textbf{k}}=\hbar c |\textbf{k}|$  of the photon spectrum
(non analytic at $\textbf{k}=0$). Massive photons $\hbar
\omega_{\textbf{k}}=c\sqrt{(\hbar |\textbf{k}|)^{2}+(mc)^{2}}$ do
not produce Casimir forces for the same reasons as for Fermions. On
the other hand, in the high temperature regime, the Casimir term
$-{\zeta(3)}/{16 \pi \beta d^{2}}$ becomes purely classical and is
entirely due to fluctuating Coulomb potentials in the conductor
walls \cite{Buenzli-Martin1} - \cite{Janco-Samaj}. Moreover, a
classical Coulomb system enclosed in a slab with perfectly
conducting walls manifests an universal finite size correction of
the Casimir type of the opposite sign ${\zeta(3)}/{16 \pi \beta
d^{2}}$
\cite{Janco-tellez}. Such Coulomb systems are critical at all temperatures
because of the  potential fluctuations  are always of a long range,
although charge correlations are themselves of a short range as a consequence of screening.\\

We note that the Casimir terms found in (b) are classical and
universal, namely they do not depend on the Planck constant and the
particle mass. In fact, for the free Bose-gas, it follows from a
simple dimensional analysis that such terms are necessarily of the
form $C\, {k_{B}T}/{d^{2}}$ where $C$ is a numerical constant. They
are present at all positive temperature provided that  the density
$\rho$ of the gas is higher than the critical density
$\rho_{c}(\beta)$. Since $M$ is arbitrary in (\ref{asympt-mu=0-Per})
and (\ref{asympt-mu=0-D-N}), we see that the Casimir amplitude is
dominant as soon as $d>\lambda$. For instance in liquid Helium at
$T=2$K, one has $\lambda\sim\rho^{-1/3}\sim 4$\AA, i. e. a slab of
thickness slightly large than the interatomic distance already shows
the Casimir effect. If we fix $d$ and let $T\to 0$ we enter in the
regime $\lambda\gg d$. By (\ref{low-activ}) it is sufficient to keep
the $n=r=1$ term which gives
\begin{equation}
\varphi_d (T,\mu)\sim
-\frac{e^{\beta r \mu} }{2\pi\beta\lambda^2}
e^{-    \pi^{2} (\lambda /d)^2/2}\,,\quad \mu\leq 0\,,\quad \lambda^{2}=\frac{\hbar^{2}}{m k_{B}T}
\label{asympt-d}
\end{equation}
So in contrast to the photon case the grand-potential is
exponentially small as $T\to 0$, which excludes Casimir effect in
this regime.

\acknowledgments
The paper was originated during V.A.Z.'s visits to Ecole
Polytechnique F\'{e}d\'{e}rale de Lausanne. He gratefully
acknowledges warm hospitality at Institute of Theoretical Physics
-EPFL.


\begin{thebibliography}{0}

\bibitem{Milonni}
\Name{Milonni P. W.}
\Book{The Quantum Vacuum}
\Publ{Academic Press}
\Year{1993}

 \bibitem{Brankov}
\Name{Brankov J. G., Danchev D. M., \and Tonchev N. S.}
\Book{Theory of Critical Finite-Size Systems}
\Publ{World Scientific}
\Year{2000}

\bibitem{Krech}
\Name{Krech M.}
\Book{Casimir Effect in Critical Systems}
\Publ{World Scientific}
\Year{1994}

\bibitem{Huang}
\Name{Huang K.}
\Book{Statistical Mechanics}
\Publ{Wiley}
\Year{1987}

\bibitem{Euler}
\Name{Abramowitz M. \and Stegun I. A.}
\Book{Hand Book of Mathematical Functions}
\Publ{Dover}
\Year{1972}

\bibitem{chatter}
\Name{Chatterji S. D. }
\Book{Cours d'analyse 2}
\Publ{Presses Polytechniques et Universitaires Romandes}
\Year{1997}

\bibitem{Jacobi}
\Name{Hunter J. \and Nachtergaele B.}
\Book{Applied Analysis}
\Publ{World Scientific}
\Year{2000}

\bibitem{ZRK}
\Name{Zandy R., Rudnick J. \and Kardar M. }
\REVIEW{Phys.
Rev. Lett.}{93}{2004}{155302-1}

\bibitem{Zif}
\Name{Ziff M., Uhlenbeck G. E. \and Kac M. }
\REVIEW{Phys. Rep.}{32}{1977}{169}

\bibitem{LP}
\Name{Lewis J.T. \and Pul\`{e} J.V. }
\REVIEW{Commun. Math. Phys.}{36}{1974}{1}

\bibitem{PZ}
\Name{ Pul\'{e} J.V.\and Zagrebnov V. A.}
\REVIEW{J. Math. Phys.}{45}{2004}{3565}

\bibitem{Buenzli-Martin1}
\Name{Buenzli P. R. \and Martin Ph. A.}
\REVIEW{J. Stat. Phys.}{119}{2005}{273}

\bibitem{Buenzli-Martin2}
\Name{Buenzli P. R. \and Martin Ph. A.}
{\it The Casimir force at high temperature}, cond-mat/0506303,
submitted to Euro. Phys. Lett.
\bibitem{Janco-Samaj}

\Name{Jancovici B. \and Samaj L.}
{\it Casimir force between two ideal-conductor walls revisited}, cond-mat/0506363,
submitted to Euro. Phys. Lett.

\bibitem{Janco-tellez}
\Name{Jancovici B. \and T\'ellez G.}
\REVIEW{J. Stat. Phys.}{82}{1996}{609}

\end{thebibliography}
\end{document}